\documentclass[aps,preprint,superscriptaddress,nofootinbib]{revtex4}

\begin{document}

\title{\bf Canonical Analysis of the First Order Form of Topologically 
Massive Electrodynamics}

\author{R. N. Ghalati}
\email{rnowbakh@uwo.ca}
\affiliation{Department of Applied Mathematics,
University of Western Ontario, London, N6A~5B7 Canada}


\author{N. Kiriushcheva}
\email{nkiriush@uwo.ca}
\affiliation{Department of Applied Mathematics,
University of Western Ontario, London, N6A~5B7 Canada}


\author{S. V. Kuzmin}
\email{skuzmin@uwo.ca}
\affiliation{Department of Applied Mathematics,
University of Western Ontario, London, N6A~5B7 Canada}
\affiliation{Huron University College,
London, N6G~1H3 Canada}
\affiliation{Physics and Astronomy Department,
University of Western Ontario, London, N6A~3K7 Canada}


\author{D. G. C. McKeon}
\email{dgmckeo2@uwo.ca}
\affiliation{Department of Applied Mathematics, 
University of Western Ontario, London, N6A~5B7 Canada}

\date{\today}

\begin{abstract}
The first order form of a three dimensional $U(1)$ gauge theory in which a 
gauge invariant mass term appears is analyzed using the Dirac procedure. 
The form of the gauge transformation which leaves the action invariant is 
derived from the constraints present.
\end{abstract}

\maketitle

\section{Introduction}

The standard Maxwell action for a $U(1)$ gauge field $A_\mu$
$$S_m = -\frac{1}{4} \int dV \left(\partial_\mu A_\nu - \partial_\nu A_\mu\right)\left(\partial^\mu A^\nu - \partial^\nu A^\mu\right)\eqno(1)$$
can be supplemented in three dimensions with a topological Chern-Simons action
$$S_{cs} = - \frac{m}{2} \int dV \epsilon^{\mu\nu\lambda} \left(\partial_\mu A_\nu - \partial_\nu A_\mu\right) A_\lambda\eqno(2)$$
so as to provide a mass to the field $A_\mu$ [1].
Both the actions $S_m$ and $S_{cs}$ are invariant under a $U(1)$ gauge transformation
$$\delta A_\mu = \partial_\mu \zeta .\eqno(3)$$

The action $S_m$ by itself can be written in first order form
$$S_m^{(1)} = \int dV \left[\frac{1}{4} F_{\mu\nu} F^{\mu\nu} - \frac{1}{2} F^{\mu\nu} \left(\partial_\mu A_\nu - \partial_\nu A_\mu\right)\right];\eqno(4)$$
here $F_{\mu\nu}$ and $A_\lambda$ are independent fields. When the equation of motion for $F_{\mu\nu}$ is used to eliminate $F_{\mu\nu}$ from eq.\  (4), one recovers $S_m$ in eq.\  (1).

The first order form of $S_m + S_{cs}$ is somewhat more involved [2],
$$S = \int d^3x \left[\frac{1}{4} W^{\mu\nu} W_{\mu\nu} - \frac{1}{2} W^{\mu\nu} \left(\partial_\mu A_\nu - \partial_\nu A_\mu \right)\right.\nonumber$$
$$\left. - \frac{m}{2} \epsilon^{\mu\nu\lambda} W_{\mu\nu} A_\lambda - \frac{m^2}{2} A^\mu A_\mu \right]. \eqno(5)$$
The equations of motion for $W_{\mu\nu}$ and $A_\mu$ are
$$W^{\mu\nu} = \partial^\mu A^\nu - \partial^\nu A^\mu + m\epsilon^{\mu\nu\lambda} A_\lambda \eqno(6)$$
and
$$\partial_\mu W^{\mu\nu} = \frac{m}{2} \epsilon^{\alpha\beta\nu} W_{\alpha\beta} +
m^2 A^\nu. \eqno(7)$$
Substitution of $W^{\mu\nu}$ from eq.\  (6) into eq.\  (5) recovers $S_m + S_{cs}$. It is apparent that the action of eq.\  (5) is invariant under the gauge transformation of eq.\  (3) provided we also transform $W^{\mu\nu}$
$$\delta W^{\mu\nu} = m \epsilon^{\mu\nu\alpha} \partial_\alpha \zeta .\eqno(8)$$

We now will demonstrate how the Dirac analysis of constrained systems [3] can be used to analyze actions which are first order in derivatives by applying this procedure to the actions of eqs.\  (4) and (5). It is of particular interest to show how this approach can be used to derive the gauge invariances of eqs.\  (3) and (8).

\section{Maxwell Electrodynamics}

The action of eq.\  (4) in four dimensional spacetime (with signature (+ + + $-$)) can be written as
$$S_m^{(1)} = \int d^4x \left[ \frac{1}{2} \left(\vec{B}^2 - \vec{E}^2\right) + \vec{E} \cdot
\left(\dot{\vec{A}} - \nabla A\right) - \vec{B}\cdot \nabla \times \vec{A}\right]\eqno(9)$$
where
$$B^i = \frac{1}{2} \epsilon^{ijk}F^{jk},\;\;\;\; E^i = F^{i0},\;\;\;\;A = A^0\;\; .\eqno(10)$$
The momenta conjugate to $\vec{B}$, $\vec{E}$, $\vec{A}$ and $A$ give rise to ten primary constraints
$$\vec{\Pi}_B = 0 ,\;\;\;\; \vec{\Pi}_E = 0 ,\;\;\;\; \vec{\pi} - \vec{E} =0 ,\;\;\;\;
\pi = 0,\eqno(11-14)$$
with eqs. (12, 13) constituting a pair of second class constraints.

The canonical Hamiltonian associated with eq.\  (9) is
$$H_c = \frac{1}{2} \left(\vec{E}^2 - \vec{B}^2\right) + \vec{E} \cdot \nabla A + \vec{B} \cdot \nabla \times \vec{A} .\eqno(15)$$
This is supplemented with Lagrange multiplier fields that ensure that the constraints are satisfied to yield the total Hamiltonian
$$H_T = H_c + \vec{\Lambda}_B \cdot \vec{\Pi}_B + \vec{\Lambda}_E  \cdot \vec{\Pi}_E +
\vec{\lambda} \cdot \left(\vec{\pi} - \vec{E}\right) + \lambda\pi .\eqno(16)$$

For consistency, the constraints must be time independent and so must have a vanishing Poisson bracket (PB) with $H_T$. We hence find that
$$\left\lbrace \vec{\Pi}_B, H_T \right\rbrace = \vec{B} - \nabla \times \vec{A}
, \eqno(17)$$
$$\left\lbrace \vec{\Pi}_E, H_T \right\rbrace = - \vec{E} - \nabla A + \vec{\lambda},\eqno(18)$$
$$\left\lbrace \vec{\pi} - \vec{E}, H_T \right\rbrace = \nabla \times \vec{B} - \vec{\Lambda}_E, \eqno(19)$$
$$\left\lbrace \vec{\pi}, H_T \right\rbrace = \nabla \cdot \vec{E}\eqno(20)$$
must vanish. From eqs.\ (18) and (19) it is apparent that the Lagrange multipliers associated with the primary second class constraints are determined, while eqs.\ (17) and (20) constitute a pair of secondary constraints. These secondary constraints must themselves have vanishing PB with $H_T$ for consistency, and so
$$\left\lbrace \vec{B} - \nabla \times \vec{A}, H_T \right\rbrace = \vec{\Lambda}_B - \nabla \times \vec{\lambda} =0,\eqno(21)$$
$$\left\lbrace \nabla \cdot \vec{E}, H_T \right\rbrace = \nabla \cdot \vec{\Lambda}_E =0,\eqno(22)$$
so that $\vec{\Lambda}_B$ is determined by eqs.\ (18) and (21) while the longitudinal part of $\vec{\Lambda}_E$ must vanish, which is consistent with eq.\  (19). The only Lagrange multiplier left undetermined is $\lambda$.

One can now either use eqs.\ (17-22) to eliminate those Lagrange multipliers which have been determined by the consistency conditions, or alternatively replace the PB by Dirac brackets (DB) and set all second class constraints equation to zero [3]. The constraints that constitute second class constraints are eqs.\ (11-13, 17, 20), and then elimination results in the following DB
$$\left\lbrace A_i\left( \vec{x}, t\right), \pi_j\left(\vec{y}, t\right) \right\rbrace^* =
\left(\delta_{ij} - \frac{\partial_i\partial_j}{\partial^2}\right)
\delta\left(\vec{x} - \vec{y}\right)\eqno(23)$$
with all other DB being identical to the corresponding PB. The form of $H_T$ is, upon elimination of the second class constraints,
$$H_T = \frac{1}{2} \left(\vec{\pi}^2 + \vec{B}^2\right) - A \nabla \cdot \vec{\pi} + \lambda\pi .\eqno(24)$$
Only the primary constraint of eq.\  (14) and the secondary constraint of eq.\  (20) are first class. The field $A$ becomes a Lagrange multiplier field. Eq.\  (24) gives the same expression for $H_T$ that one obtains if the second order form for $S_m$ in eq.\  (1) is treated using the Dirac procedure [3].

Two approaches to determine the gauge transformation that leaves the action invariant are provided in refs.\ [4, 5], with the latter reference providing a somewhat more general procedure. In both cases we find that the gauge generator is
$$G(\zeta, \dot{\zeta}) = -\int d^3x \left(\zeta \nabla \cdot \vec{\pi} + \dot{\zeta}\pi\right)\eqno(25)$$
so that
$$\delta A = \left\lbrace A, G\right\rbrace^* = -\dot{\zeta}\eqno(26)$$
and
$$\delta\vec{A} = \left\lbrace \vec{A}, G\right\rbrace^* = \nabla \zeta ,\eqno(27)$$
while
$$\delta\vec{E} = \left\lbrace \vec{E}, G\right\rbrace^* = 0 = \delta\vec{B} .\eqno(28)$$
Together then, $\delta A_\mu = \partial_\mu\zeta$ and $\delta F_{\mu\nu} = 0$, as one would expect from inspection of $S_m^{(1)}$ in eq.\  (4).

We now apply the Dirac formalism to the more interesting (and complicated) case of the action $S$ of eq.\  (5).

\section{Topologically Massive Electrodynamics}

The action of eq.\  (5) can be written as
$$S = \int d^3x \left[ \frac{1}{2} \left( W^2 - \vec{W}^2\right) + \frac{m^2}{2}
\left( A^2 - \vec{A}^2\right) - \left(W\nabla \times \vec{A} + \vec{W} \cdot \dot{\vec{A}} +
\vec{W} \cdot \nabla A\right) \right. \nonumber$$ 
$$\left. - m \left(\vec{W} \times \vec{A} + WA\right) \right]\eqno(29)$$
if the metric has diagonal $(+ + -)$, $\epsilon_{012} = 1$, $A = A^0$, $W = \frac{1}{2} \epsilon_{ij} W^{ij}$, $W^i = W^{0i}$ and $\vec{U} \times \vec{V} = \epsilon_{ij}U^iV^j$. The momenta associated with $A$, $\vec{A}$, $W$ and $\vec{W}$ are now given by the primary constraints
$$\pi = 0,\;\;\;\vec{\pi} + \vec{W} = 0,\;\;\;\Pi = 0, \;\;\;\vec{\Pi} = 0\eqno(30-33)$$
respectively. The constraints of eqs.\ (31, 33) are second class; if DB are used it is possible to immediately replace $\vec{W}$ by $-\vec{\pi}$ in the canonical Hamiltonian and we obtain
$$H_c = \frac{1}{2} \left( \vec{\pi}^2 - W^2\right) + \frac{m^2}{2} \left( \vec{A}^2 - A^2\right)
+ W \nabla \times \vec{A} - A \nabla \cdot \vec{\pi} + m \left(AW - \vec{\pi} \times \vec{A}\right) .\eqno(34)$$

Consistency means that $\dot{\Pi} = \left\lbrace \Pi , H_c \right\rbrace $ should vanish; with $H_c$ given by eq.\  (34) then
$$\left\lbrace \Pi , H_c\right\rbrace = W - \nabla \times \vec{A} - m A = 0 \eqno(35)$$
is a secondary constraint. Similarly, as $\dot{\pi} = 0$, we obtain another secondary constraint
$$\left\lbrace \pi , H_c\right\rbrace = m^2 A + \nabla \cdot \vec{\pi} - mW = 0
. \eqno(36)$$

Eqs.\ (30, 32, 35, 36) together form four constraints. However, the PB of these four constraints form a matrix with rank two; consequently appropriate linear combinations of these four constraints can be chosen so that two are first class and two are second class. A suitable pair of first class constraints are
$$\gamma_1 = \pi + m\Pi, \;\;\;\;\;\;\;\gamma_2 = m \nabla \times \vec{A} - \nabla \cdot \vec{\pi}\eqno(37,38)$$
and of second class constraints are (provided $m \neq 0$)
$$\chi_1 = \Pi,\;\;\;\;\;\;\;\chi_2 = \nabla \times \vec{A}+ m A - W .\eqno(39,40)$$
Using eq.\  (40) to eliminate $W$ in eq.\  (34) leads to
$$H_c = \frac{1}{2} \left[ \vec{\pi}^2 + \left(\nabla \times \vec{A}\right)^2 + m^2\vec{A}^2\right] + A\left(m \nabla \times \vec{A} - \nabla \cdot \vec{\pi}\right) + m \vec{\pi} \times \vec{A} .\eqno(41)$$
It is now evident that
$$\left\lbrace \gamma_1 , H_c\right\rbrace = -\gamma_2, \;\;\;\;
\left\lbrace \gamma_2 , H_c\right\rbrace = 0, \;\;\;\;
\left\lbrace \gamma_1 , \gamma_2\right\rbrace = 0 .\eqno(42-44)$$
With the first class constraints of eqs.\ (37, 38) satisfying the commutation relations of eqs.\ (42-44), the methods of refs. [4, 5] lead to the gauge generator
$$G = \int d^2x \left[ - \dot{\zeta} \pi + \zeta (m \nabla \times \vec{A} - \nabla \cdot \vec{\pi})\right]\eqno(45)$$
so
$$\delta A = \left\lbrace A, G\right\rbrace = - \dot{\zeta} \eqno(46)$$
and
$$\delta \vec{A} = \left\lbrace \vec{A}, G\right\rbrace = \nabla \zeta , \eqno(47)$$
as well as
$$\delta \pi_i = -m\epsilon_{ij} \partial_j\zeta .\eqno(48)$$
Eqs.\ (31) and (40) can now be used to show that
$$\delta W = -m\dot{\zeta},\eqno(49)$$
$$\delta W_i = m\epsilon_{ij} \partial_j \zeta .\eqno(50)$$
Together, from eqs.\ (46, 47, 49, 50) we recover the gauge transformations of eqs.\ (3, 8).

\section{Discussion}

The canonical structure of the first order form of the Maxwell and Maxwell plus Chern-Simons actions have been analyzed in some detail here. Our interest in these particular models was stimulated by ref. [6]. However, more importantly, the procedure outlined should serve as a model for how to perform a fully consistent canonical analysis of the Einstein-Hilbert (EH) action in general relativity when it is expressed in first order form. In refs. [7-10], the Lagrangian $\sqrt{-g} R(\Gamma)$, a first order action which is inequivalent to the second order Einstein-Hilbert action when in two dimensions,\footnote{In two dimensions, the first order Lagrangian $\sqrt{-g} R(\Gamma)$ is not equivalent to the second order Einstein-Hilbert action $\sqrt{-g} R(g)$, as the equation of motion for the affine connection $\Gamma$ does not yield the Christoffel symbol.  However, in dimensions greater than two, these two actions are in fact equivalent.} is analyzed using the Dirac constraint formalism employed above. As has been noted in refs. [9, 10], the usual Dirac-Arnowitt-Deser-Misner approach [11, 12] to the canonical structure of the EH action involves elimination at the outset of canonical variables through use of all equations of motion that are independent of time derivatives, irrespective of whether these equations correspond to first or second class constraints. (This is most explicitly seen in the presentation appearing in ref. [13].) Using first class constraints to eliminate canonical variables can be seen from the above examples to destroy whatever gauge symmetry is generated by these first class constraints. We hope to circumvent this shortcoming in the analysis of refs. [11, 12] through a careful application of the Dirac constraint formalism (in which first class constraints are not used to eliminate dynamical degrees of freedom) to the first order EH action in dimensions higher than two.

\section{Acknowledgements}

NSERC provided financial support to D. G. C. McKeon. Roger Macloud had a helpful suggestion.

\end{document}